\documentclass[review]{elsarticle}

\usepackage{graphicx}
\usepackage{dcolumn}
\usepackage{bm}
\usepackage{CJK}
\usepackage{url}
\usepackage{color}

\begin{document}


\title{Symmetry energy of super-dense neutron-rich matter from integrating barotropic pressures in neutron stars and heavy-ion reactions}


\author{Bao-An Li$^{1}$\footnote{Corresponding author: Bao-An.Li@Tamuc.edu} and Wen-Jie Xie$^{1,2}$}
\address {$^{1}$Department of Physics and Astronomy, Texas A\&M University-Commerce, TX 75429-3011, USA}
\address {$^{2}$Department of Physics, Yuncheng University, Yuncheng 044000, China}

\date{\today}

\begin{abstract}
Within the minimum model of neutron stars (NS) consisting of neutrons, protons and electrons, a new approach is proposed for inferring the symmetry energy of super-dense neutron-rich nucleonic matter above twice the saturation density $\rho_0$ of nuclear matter directly from integrating iteratively barotropic pressures in both neutron stars and heavy-ion reactions. Simultaneously, the proton fraction of NSs at $\beta$ equilibrium is extracted as a function of baryon density from the same procedure. An application of this approach using the NS pressure from GW170817 and the pressure in cold symmetric nuclear matter (SNM) extracted earlier by analyzing nuclear collective flow data in relativistic heavy-ion collisions provides a useful constraining band for the symmetry energy above $2\rho_0$. 
\end{abstract}

\maketitle
\newpage
\noindent{\bf Motivation:} A longstanding and shared goal of many astrophysical observations and terrestrial nuclear experiments is to understand the Equation of State (EOS) of dense neutron-rich nuclear matter. Indeed, much efforts have been devoted to extracting information about the EOS in both fields using various observatories and facilities during the last few decades. On one hand, by analyzing X-rays from individual neutron stars (NSs) and/or gravitational waves from their mergers, the pressure $P_{\rm{NS}}$ as a function of energy density $\epsilon$ or baryon density $\rho$ has often been extracted. However, generally no information about the composition of the source (e.g. the proton fraction in NSs) is extracted simultaneously. This is partially because normally isospin-independent polytropic EOSs are used in the analyses and a barotropic pressure is sufficient for solving the  Tolman-Oppenheimer-Volkov (TOV) equations \cite{Tolman34,Oppenheimer39}. For example, the blue band shown in Fig.1 is the $P_{\rm{NS}}(\rho)$ as a function of baryon density at 90\% confidence level in canonical NSs at $\beta$ equilibrium reported by the LIGO \& VIRGO Collaborations from their analyses of the GW170817 event \cite{LIGO2018}. On the other hand, systematic analyses of terrestrial nuclear experiments over the last few decades have set a reasonably tight bound on the EOS of cold, symmetric nuclear matter (SNM) from sub-saturation densities to about four times the saturation density $\rho_0$ of nuclear matter \cite{Dan02,Trau12,Garg18,Wolfgang19}. As an example, shown in Fig.1 with the red band is the pressure $P_{\rm{SNM}}(\rho)$ in SNM extracted from transport model analyses of nuclear collective flow \cite{Dan02} in heavy-ion collisions at beam energies from about 0.4 to 10 GeV/nucleon. Since there are confusions and incorrect statements in the literature about what EOS information can be extracted from heavy-ion reactions, we emphasize here that while heavy-ion reactions create hot matter, the underlying EOS of cold matter used as a basic input of transport model simulations of these reactions can be reliably extracted from comparing the simulations with experimental observables, see, e.g., refs. \cite{Dan02,Bert,Sto86,Cas90,Bass,Li2008,Xu19}. Of course, there are still many uncertainties involved in extracting the cold EOS from heavy-ion reactions. However, to our best knowledge, currently they are not more than those involved in extracting the EOS from gravitational waves emitted by merging NSs. In fact, both the $P_{\rm{NS}}(\rho)$ and $P_{\rm{SNM}}(\rho)$ shown in Fig. 1 have been used extensively in many different ways to constrain various aspects of nuclear many-body theories and interactions in the literature, see, e.g., refs. \cite{EPJA-review, JPG-review,CUSTIPEN}. While both pressures still have rather large error bands, to  our best knowledge, they represent the state of the art of the  fields.

 \begin{figure}[htb]
\begin{center}
\resizebox{0.9\textwidth}{!}{
  \includegraphics{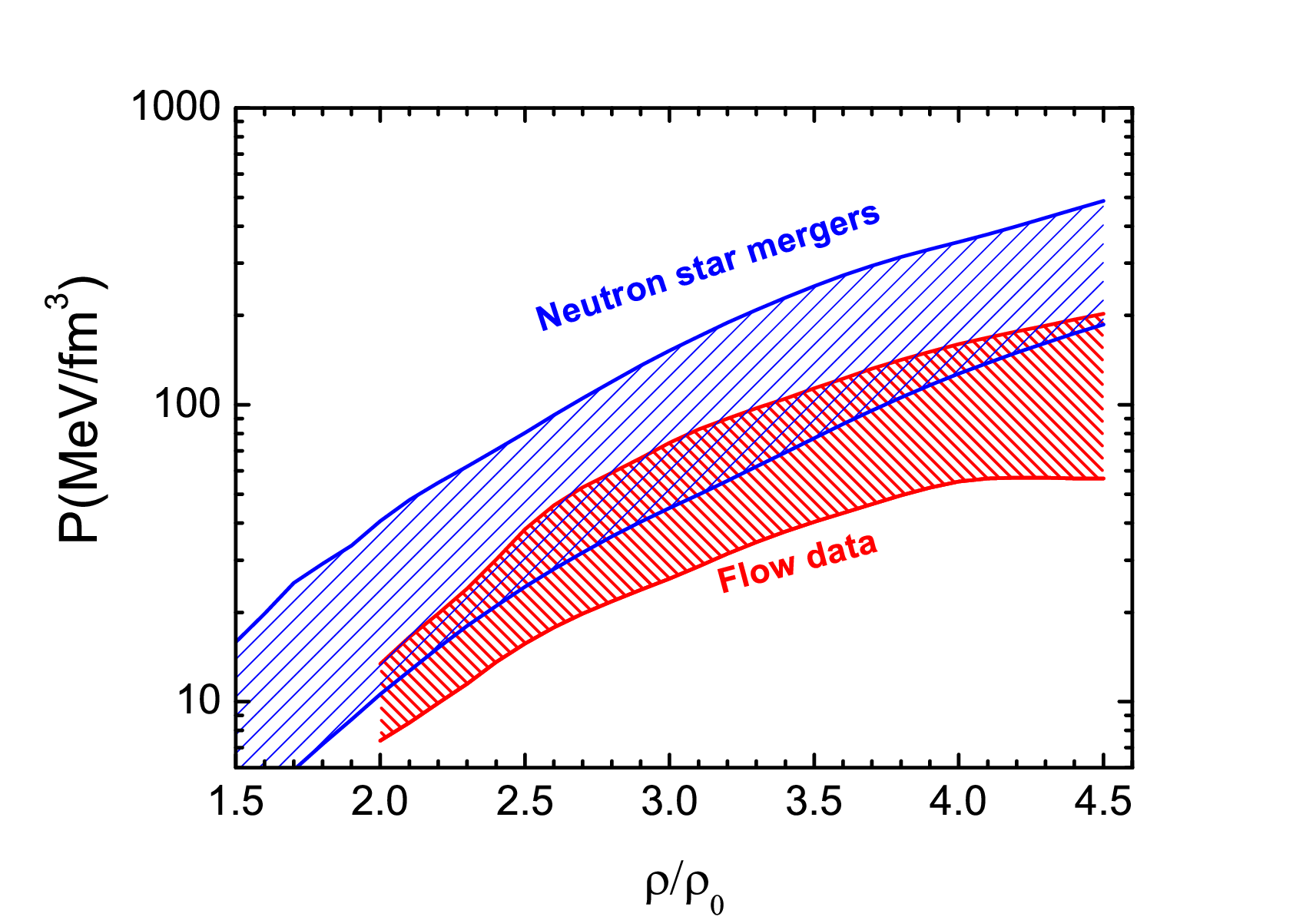}
  }
\caption{(Color online) The pressure as a function of baryon density in neutron stars (blue) extracted by the LIGO \& VIRGO Collaborations from their analyses of the GW170817 event \cite{LIGO2018}. The red band is the pressure in cold symmetric nuclear matter extracted from analyzing nuclear collective flow \cite{Dan02} in heavy-ion collisions with beam energies from about 0.4 to 10 GeV/nucleon.}\label{figure1}
\end{center}
\end{figure}

It has been broadly recognized that the high-density behavior of nuclear symmetry energy $E_{\rm{sym}}(\rho )$ is the most uncertain aspect of the EOS of dense neutron-rich nucleonic matter besides possible phase transitions \cite{Li2008,ditoro,Steiner05,Lat12,Tsang12,Chuck14,Tesym,Bal16,Oertel17,Chen2017,Li17,Blas18,MM1,PPNP-Li,Baiotti}.
Indeed, the pressure $P_{\rm{NS}}(\rho)$ from GW170817 has already been used to constrain some features of the $E_{\rm{sym}}(\rho )$ within several approaches, see, e.g. refs. \cite{NBZ19a,Zhang19apj,Tsang19}. Since information about the composition of NSs is rarely extracted from analyses of astrophysical data so far, one often assumes that NSs are made of pure neutron matter (PNM) in many studies for various purposes. Such an extreme assumption implies that the $E_{\rm{sym}}(\rho )$ is zero in NSs at $\beta$ equilibrium. It is intrinsically inconsistent with the stated goal of extracting the underlying symmetry energy which determines uniquely the proton fraction $x_p(\rho)$ in NSs at $\beta$equilibrium (see, e.g., Eq. (\ref{xp})). Moreover, information about the  density dependence of $x_p(\rho)$ has many important ramifications on properties of NSs especially the cooling mechanisms of protoneutron stars. So, is there any straightforward and self-consistent way to infer the high-density $E_{\rm{sym}}(\rho )$ directly from the $P_{\rm{NS}}(\rho)$ and $P_{\rm{SNM}}(\rho)$ shown in Fig.1 without having to assume NSs are made of purely neutrons? Yes, there is. Here we propose a novel approach and present numerical results of its first application, providing a useful constraining band for the $E_{\rm{sym}}(\rho )$ at $\rho \ge 2.0\rho_0$ where it is still observationally/experimentally largely unknown. Moreover, we extract simultaneously the proton fraction $x_p(\rho)$ in the same density range in NSs at $\beta$ equilibrium.\\

\noindent{\bf Approach:} It is well known that the EOS of cold asymmetric nucleonic matter (ANM) of isospin asymmetry $\delta$ and density $\rho$ can be
approximated as \cite{Bom91}
\begin{equation}\label{eos1}
E(\rho ,\delta )=E_{\rm{SNM}}(\rho)+E_{\rm{sym}}(\rho )\delta ^{2} +\mathcal{O}(\delta^4)
\end{equation}
in terms of the energy per nucleon $E_{\rm{SNM}}(\rho)\equiv E(\rho ,\delta=0)$ in SNM and  the nuclear  symmetry energy $E_{\rm{sym}}(\rho )$. The corresponding pressure in ANM can then be written as
\begin{equation}\label{pressure}
  P(\rho, \delta)=\rho^2\frac{dE(\rho,\delta)}{d\rho}=\rho^2[\frac{dE_{\rm{SNM}}(\rho)}{d\rho}+\frac{dE_{\rm{sym}}(\rho)}{d\rho}\delta^2]=P_{\rm{SNM}}(\rho)+\rho^2\frac{dE_{\rm{sym}}(\rho)}{d\rho}\delta^2
\end{equation}
where $P_{\rm{SNM}}(\rho)\equiv \rho^2\frac{dE_{\rm{SNM}}(\rho)}{d\rho}$ is the pressure in SNM while the pressure in PNM $P_{\rm{PNM}}(\rho)\equiv P(\rho,\delta=1)$ can be written as
\begin{equation}\label{pressure}
  P_{\rm{PNM}}(\rho)=P_{\rm{SNM}}(\rho)+\rho^2\frac{dE_{\rm{sym}}(\rho)}{d\rho}.
\end{equation}
Therefore, if both the $P_{\rm{SNM}}(\rho)$ and $P_{\rm{PNM}}(\rho)$ are known from terrestrial nuclear experiments and/or astrophysical observations, the above equation can then be easily inverted to obtain the symmetry energy $E_{\rm{sym}}(\rho )$ via
\begin{equation}\label{esym}
  E_{\textrm{sym}}(\rho)=E_{\textrm{sym}}(\rho_i)+\int_{\rho_i}^{\rho}\frac{P_{\rm{PNM}}(\rho_v)-P_{\rm{SNM}}(\rho_v)}{\rho_v^2}d\rho_v
\end{equation}
where $\rho_i$ is a reference density below which the $E_{\rm{sym}}(\rho )$ is known.
To apply the Eq. (\ref{esym}), one needs the density dependence of both the pressure $P_{\rm{SNM}}(\rho)$ in SNM and the pressure $P_{\rm{PNM}}(\rho)$ in PNM. While the $P_{\rm{SNM}}(\rho)$ has been extracted explicitly from analyzing heavy-ion reactions using mostly transport models where the EOS of cold SNM is a basic input \cite{Dan02,Bert}, the pressure $P_{\rm{PNM}}(\rho)$ in PNM is not the same as the NS pressure $P_{\rm{NS}}(\rho)$ extracted directly from studying properties of NSs unless one explicitly assumes that NSs are completely made of just neutrons. In fact, it is well known that the proton fraction $x_p(\rho)$ at a given density in NSs at $\beta$ equilibrium is determined uniquely by the symmetry energy $E_{\rm sym}(\rho)$. Thus, the pressure extracted from analyzing observations of NSs without knowing explicitly their compositions should not be simply used as the $P_{\rm{PNM}}(\rho)$ in extracting the underlying density dependence of nuclear symmetry energy $E_{\rm sym}(\rho)$ in whatever approaches one may choose. 

Before applying the Eq. (\ref{esym}), some more elaborations about its validity are necessary. The Eq. (\ref{esym}) is derived directly from the pressure of PNM of Eq. (\ref{pressure}) without making any additional assumption. As outlined above, we started from the well established parabolic approximation Eq. (\ref{eos1}) for the EOS of ANM in which the density $\rho$ and isospin asymmetry $\delta$ are two independent variables. By considering the pressure in the extreme case of PNM where the isospin asymmetry is a constant one, we are able to express precisely without missing anything within the parabolic approximation for the EOS of ANM the symmetry energy as an integral of the difference between $P_{\rm{PNM}}(\rho)$ and $P_{\rm{SNM}}(\rho)$.  The Eq. (\ref{esym}) is valid independent of the way one may use to get the latter two pressures as functions of density. As we mentioned earlier, while the $P_{\rm{SNM}}(\rho)$ is extracted directly from heavy-ion collisions, the extraction of $P_{\rm{PNM}}(\rho)$ from the pressure $P_{\rm{NS}}(\rho)$ in neutron stars at $\beta$ equilibrium is not straightforward as we shall discuss in detail next. In particular, the $\beta$ equilibrium and charge neutrality conditions introduce a dependence between the density $\rho$ and isospin asymmetry $\delta$, namely, a function $\delta(\rho)$. However, the latter does not affect the validity of the Eq. (\ref{esym}) as only the pressure of PNM inferred from the pressure $P_{\rm{NS}}(\rho)$ is used in Eq. (\ref{esym}). Of course, the challenge is to infer the PNM pressure $P_{\rm{PNM}}(\rho)$ from the neutron star pressure $P_{\rm{NS}}(\rho)$ reported by the LIGO/VIRGO Collaborations. Next, we present our inversion technique to meet this challenge. 

In the minimal model of charge neutral neutron stars consisting of neutrons, protons and electrons ($npe$ matter) at $\beta$ equilibrium,  
the pressure is given by \cite{Lat01}
\begin{eqnarray}\label{pre}
P_{\rm{NS}}(\rho,\delta)&=&\rho^2\left(\frac{\partial E(\rho,\delta)}{\partial \rho}
\right)_{\delta}+\frac{1}{4}\rho_e\mu_e\nonumber\\
&=&\rho^2[\frac{dE_{\rm{SNM}}(\rho)}{d\rho}+\frac{dE_{\rm{sym}}(\rho)}{d\rho}\delta^2]
+\frac{1}{2}\delta(1-\delta)\rho E_{\rm sym}(\rho)
\end{eqnarray}
where $\rho_e=\frac{1}{2}(1-\delta)\rho$ and
$\mu_e=\mu_n-\mu_p=4\delta E_{\rm sym}(\rho)$ are, respectively,
the density and chemical potential of electrons. The value of the
isospin asymmetry $\delta$ (or the corresponding proton fraction $x_p=(1-\delta)/2$)
at $\beta$ equilibrium is determined completely by the symmetry energy through the 
the chemical equilibrium and charge neutrality conditions. The resulting $x_p(\rho)$ can be written as \cite{Lat01}
\begin{eqnarray}\label{xp}
x_p(\rho)= 0.048 \left[E_{\rm sym}(\rho)/E_{\rm sym}(\rho_0)\right]^3
(\rho/\rho_0)(1-2x_p(\rho))^3.
\end{eqnarray}
We emphasize that the NS pressure $P_{\rm{NS}}(\rho,\delta)$ becomes barotropic, i.e., depending only on the density, once the density profile of the proton fraction $x_p(\rho)$ (and the corresponding $\delta(\rho)$) at $\beta$ equilibrium is determined. In the following discussions about how to find the right $x_p(\rho)$ (or $\delta(\rho)$) through an iteration procedure, we keep using the notation $P_{\rm{NS}}(\rho,\delta)$ instead of the final $P_{\rm{NS}}(\rho)$ for the model NS pressure. 

To see how one may use progressively more accurately the available pressure $P_{\rm{NS}}(\rho)$ extracted from observations without any NS composition information in evaluating the $E_{\rm sym}(\rho)$, it is useful to first recast the NS pressure in Eq. (\ref{pre}) in terms of the $P_{\rm{PNM}}(\rho)$, $P_{\rm{SNM}}(\rho)$ and $x_p(\rho)$ as
\begin{equation}
P_{\rm{NS}}(\rho,\delta)=P_{\rm{SNM}}(\rho)+\rho^2(1-2x_p)^2\frac{dE_{\rm{sym}}(\rho)}{d\rho}+\rho x_p(1-2x_p)E_{\rm{sym}}(\rho).
\end{equation}
After expanding the second term $(1-2x_p)^2$ and using the definition of Eq. (\ref{pressure}) for $P_{\rm{PNM}}(\rho)$, the above equation can be rewritten as
\begin{eqnarray}
P_{\rm{NS}}(\rho,\delta)=P_{\rm{PNM}}(\rho)-x_p\rho[4\rho\frac{dE_{\rm{sym}}(\rho)}{d\rho}-E_{\rm{sym}}(\rho)]
+2x_p^2\rho[2\rho\frac{dE_{\rm{sym}}(\rho)}{d\rho}-E_{\rm{sym}}(\rho)].
\end{eqnarray}
Thus, to calculate the $E_{\rm sym}(\rho)$ using the Eq. (\ref{esym}) one may use  as input
\begin{equation}\label{ppnm}
P_{\rm{PNM}}(\rho)=P_{\rm{NS}}(\rho)+x_p\rho[4\rho\frac{dE_{\rm{sym}}(\rho)}{d\rho}-E_{\rm{sym}}(\rho)]
-2x_p^2\rho[2\rho\frac{dE_{\rm{sym}}(\rho)}{d\rho}-E_{\rm{sym}}(\rho)].
\end{equation}
Obviously, at the PNM limit, $x_p(\rho)=0$, one has $P_{\rm{PNM}}(\rho)=P_{\rm{NS}}(\rho)$. The last two terms proportional to $x_p(\rho)$ and $x^2_p(\rho)$ in Eq. (\ref{ppnm}) are progressively smaller quantities as $x_p(\rho)$ is always smaller than 0.5. Considering the coupled Eqs. (\ref{esym}), (\ref{xp}) and (\ref{ppnm}) together, it is seen that they constitute a closed iteration process for calculating directly the $E_{\rm sym}(\rho)$ and the corresponding $x_p(\rho)$ simultaneously using the pressure $P_{\rm{NS}}(\rho)$ from analyzing NS observations and the $P_{\rm{SNM}}(\rho)$ extracted from heavy-ion reactions. Namely, first setting $x_p(\rho)=0$ (lowest accuracy) and inserting the resulting relation $P_{\rm{PNM}}(\rho)=P_{\rm{NS}}(\rho)$ in Eq. (\ref{esym}), we calculate the symmetry energy $E_{\rm sym}(\rho)$ at the zeroth order in $x_p$. The resulting $E_{\rm sym}(\rho)$  is then used to calculate a new density profile $x_p(\rho)$ and the corresponding $\delta(\rho)$ at $\beta$ equilibrium using Eq. (\ref{xp}). Then, using the Eq. (\ref{ppnm}) one can calculate a new $P_{\rm{PNM}}(\rho)$ to be used in the Eq. (\ref{esym}) to recalculate the symmetry energy $E_{\rm sym}(\rho)$. The latter is then put back to the Eq. (\ref{xp}) to repeat the process until both the $E_{\rm sym}(\rho)$ and $x_p(\rho)$ do not change anymore, namely, until the input and output $E_{\rm sym}(\rho)$ on the two sides of Eq. (\ref{esym}) become and stay the same.\\

\noindent{\bf Inputs for the first application:} 
As the first application of the procedure detailed above, we evaluate the upper and lower limits of the $E_{\rm sym}(\rho)$ and $x_p(\rho)$ using the two pressures shown in Fig.\ 1. Since the Eq. (\ref{esym}) for calculating the $E_{\rm sym}(\rho)$ involves the difference of $P_{\rm{NS}}(\rho)$ and $P_{\rm{SNM}}(\rho)$, we calculate the upper (lower) bounds of the $E_{\rm sym}(\rho)$ and $x_p(\rho)$ by using the upper (lower) limit of $P_{\rm{NS}}(\rho)$ from GW170817 {\it but} the lower (upper) limit of $P_{\rm{SNM}}(\rho)$ from heavy-ion collisions. Naturally, this leads to the most conservative upper and lower limits of the $E_{\rm sym}(\rho)$ and $x_p(\rho)$. We note that there are some ambiguities in assigning a statistical significance to the obtained boundaries because the $P_{\rm{NS}}(\rho)$ from GW170817 is at 90\% confidence level while the $P_{\rm{SNM}}(\rho)$ from heavy-ion reactions is given in terms of its absolute (100\%) upper and lower boundaries. In addition, the resulting bands are expected to be wide because of the large uncertainties of the $P_{\rm{NS}}(\rho)$ and $P_{\rm{SNM}}(\rho)$ shown in Fig. 1. Nevertheless, we feel it is reasonable to consider the results as at 90\% confidence level. Moreover, as we shall show next, the results are still very useful compared to not only the extreme PNM model for NSs but also the $E_{\rm sym}(\rho)$ extracted from the recent Bayesian analyses using the currently available radius data of canonical NSs \cite{Xie19}. 

To infer the $E_{\rm{sym}}(\rho)$ at high densities using Eq. (\ref{esym}), we need the value of $E_{\rm{sym}}(\rho_i)$ at the reference density $\rho_i$ where the integration starts. In this work, since the $P_{\rm{SNM}}(\rho)$ from analyzing nuclear collective flow data are only given in the range of $2\rho_0$ to $4.5\rho_0$ \cite{Dan02} although there are reports of $P_{\rm{SNM}}(\rho)$ at lower densities but with different uncertainties from analyzing kaon productions in heavy-ion collisions \cite{Lynch09}, we start the integration at $2\rho_0$ as our interest here is mostly providing a guide for the $E_{\rm{sym}}(\rho)$ at super-high densities. Moreover, we take the central value of $E_{\rm{sym}}(2\rho_0)$=55.7 MeV from the ASY-EOS $E_{\rm{sym}}(\rho)$ band (shown in red in Fig. 4) from analyzing relative flows and yield ratios of light mirror nuclei in heavy-ion collisions at SIS/GSI \cite{Wolfgang19,russ11,ASY-EOS}.\\

\begin{figure}[htb]
\begin{center}
\vspace{-2.cm}
\resizebox{0.8\textwidth}{!}{
  \includegraphics{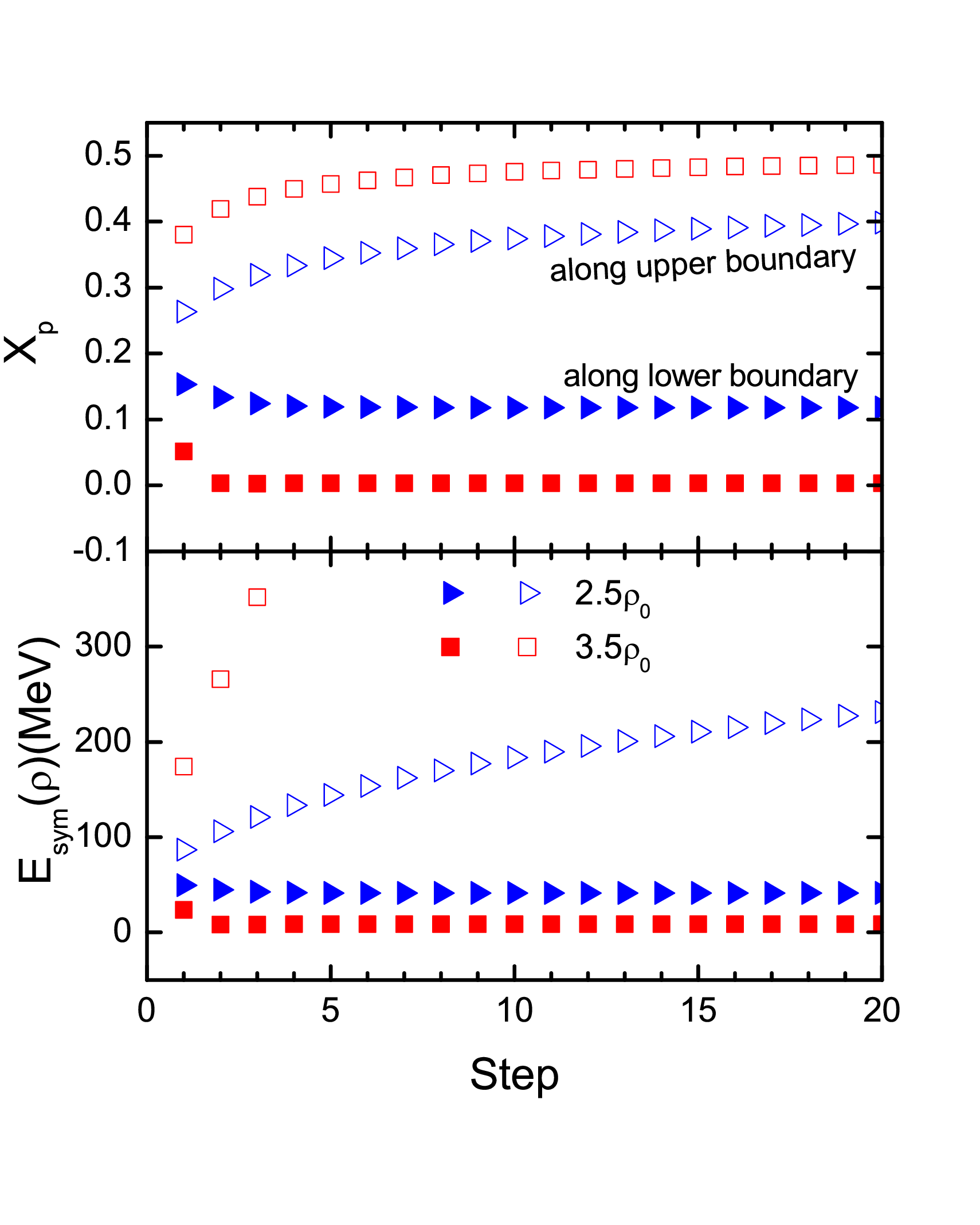}
  }
  \vspace{-1.cm}
 \caption{(Color online) The proton fraction (upper) in NSs at $\beta$ equilibrium and symmetry energy (lower) at $2.5\rho_0$ and $3.5\rho_0$ on their respective upper and lower boundaries as functions of the step number during their iterative evaluation process.}\label{figure2}
\end{center}
\end{figure}
\noindent{\bf Results and Discussions:} 
First of all, it is necessary to examine if the self-consistent iteration procedure actually works or not. For this purposes, we studied the $E_{\rm sym}(\rho)$ and $x_p(\rho)$ along their upper and lower boundaries as functions of density and iteration number from the beginning to 300 steps. As expected, they both saturate quickly especially at relatively low densities and along the lower boundary. As examples, shown in Fig. 2 are the $E_{\rm sym}(\rho)$ and $x_p(\rho)$ as functions of the step number at $2.5\rho_0$ and $3.5\rho_0$ along the two boundaries. It is seen that the $x_p(\rho)$ becomes stable after only about 5 steps for most cases while the $E_{\rm sym}(\rho)$ takes a few more steps especially at high densities along the upper boundary. Generally speaking, because of the $E_{\rm sym}(\rho)\delta^2$ term in the EOS of ANM, a higher $E_{\rm sym}(\rho)$ requires a lower $\delta$ (or higher $x_p$) simply from the energetics while their exact relation is given by Eq. (\ref{xp}) as we discussed earlier. The $x_p$ reaches its upper limit of 0.5 quickly when the $E_{\rm sym}(\rho)$ increases with density along the upper boundary. As a cubic root of the Eq. (\ref{xp}), the $x_p$ does not increases further once it has reached its upper limit even as the $E_{\rm sym}(\rho)$ is still increasing, namely the NS becomes SNM with electrons when the $E_{\rm sym}(\rho)$ becomes super-high. On the contrary, when the $E_{\rm sym}(\rho)$ becomes super-low or even negative along the lower boundary, the NS becomes PNM. These features are all expected. As indicated by the parabolic approximation of ANM EOS of Eq. (\ref{eos1}), the nuclear symmetry energy is essentially the energy cost of converting all protons in SNM into neutrons to transform the system into PNM. When the $E_{\rm sym}(\rho)$ is too high, this will not happen because it is not energetically favorible. Instead, the system becomes SNM with equal numbers of protons, neutrons and electrons even when we artificially initialized the system as a PNM. On the other hand, if the $E_{\rm sym}(\rho)$ is low, then it is easier to convert more protons into neutrons. When the $E_{\rm sym}(\rho)$ is super-low or negative, the NS quickly becomes PNM. These features from our numerical calculations clearly demonstrate that our iteration procedure worked successfully as it is supposed to.  

\begin{figure}[htb]
\begin{center}
\resizebox{0.8\textwidth}{!}{
  \includegraphics{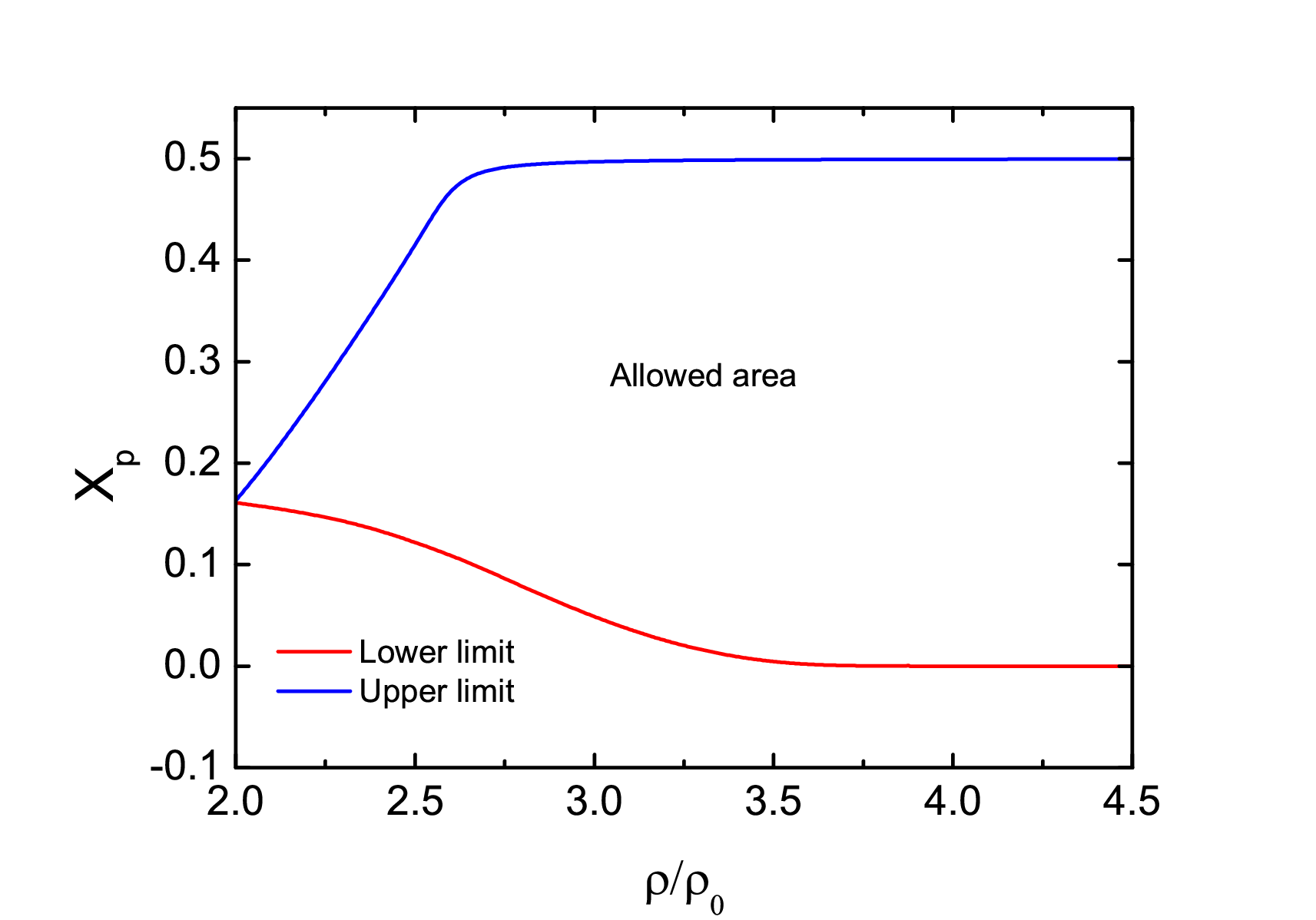}
  }
\caption{(Color online)The stabilized proton fraction $x_p(\rho)$ values along its upper and lower boundaries taken at the step 300 as functions of density.}\label{figure3}
\end{center}
\end{figure}
The stabilized $x_p(\rho)$ values along both the upper and lower boundaries taken at the step 300 are shown as functions of density in Fig. 3.  Below about $3.5\rho_0$, the  $x_p(\rho)$ band is useful for better understanding properties of NSs. For example, there are at least about 10\% protons at $2.5\rho_0$ in NSs at $\beta$ equilibrium.
As we mentioned earlier, the band is expected to be wide given the uncertainties of the input pressures we used. Nevertheless, in our opinion, even getting the limited NS composition information from combining the barotropic pressures from both neutron stars and heavy-ion reactions within the novel approach proposed here represents a significant progress in the field. At higher densities, however, the proton fraction inferred is between zero and 0.5, namely, the data we used does not constrain at all the proton fraction of NS matter at densities above about $3.5\rho_0$. Thus, narrowing down the error bands of the pressures extracted in both astrophysical observations and terrestrial nuclear experiments in the ultra-dense region will be very useful.

Now we turn to the inferred nuclear symmetry energy in the super-dense region. For comparisons and completeness, it is first necessary to comment on the current status of constraining the $E_{\rm sym}(\rho)$ at supra-saturation densities.  As discussed extensively in the literature, see, e.g., refs. \cite{ditoro,Steiner05,Li2008,Lat12,Tsang12,Chuck14,Tesym,Bal16,Oertel17,Chen2017,Li17,Blas18,PPNP-Li} and ref. \cite{Li2019review} for the latest review, the $E_{\rm{sym}}(\rho)$ has been relatively well constrained up to about $1.3\rho_0$ using various terrestrial laboratory data, while in the region of $1.3\rho_0-2.0\rho_0$
 there are still some outstanding uncertainties and controversies \cite{Wolfgang19,Li2019review,XiaoPRL}. On the other hand, the latest Bayesian analysis of the radii of canonical NSs indicates that the symmetry energy at twice the saturation density $E_{\mathrm{sym}}(2\rho_0)$ =39.2$_{-8.2}^{+12.1}$ MeV at 68\% confidence level \cite{Xie19}. This value is in very good agreement with findings of several other studies of neutron star radii, tidal deformability and maximum masses using different approaches within their statistical errors \cite{NBZ18,NBZ19a,Zhang19apj,Nakazato19,LWChen19,PKU-Meng,Baillot19,Zhou19}. Interestingly, as shown in Fig.4, the value of $E_{\mathrm{sym}}(2\rho_0)$ extracted from analyzing astrophysical data is also in reasonably good agreement with the ASY-EOS result (shown as the red bands in Fig. 4) \cite{russ11,ASY-EOS}. However, above twice the saturation density, even the trend of the $E_{\mathrm{sym}}(\rho)$ is not so clear. But it is encouraging to note that it has been found consistently in several very recent studies \cite{Zhang19apj,Xie19,Zhou19} that the very recently reported mass $M=2.14^{+0.10}_{-0.09}$~M$_\odot$ of PSR~J0740+6620 \cite{M217} can tighten significantly the lower boundary of $E_{\mathrm{sym}}(\rho)$ above about $2.5\rho_0$. 
 
\begin{figure}[htb]
\begin{center}
\resizebox{0.8\textwidth}{!}{
  \includegraphics{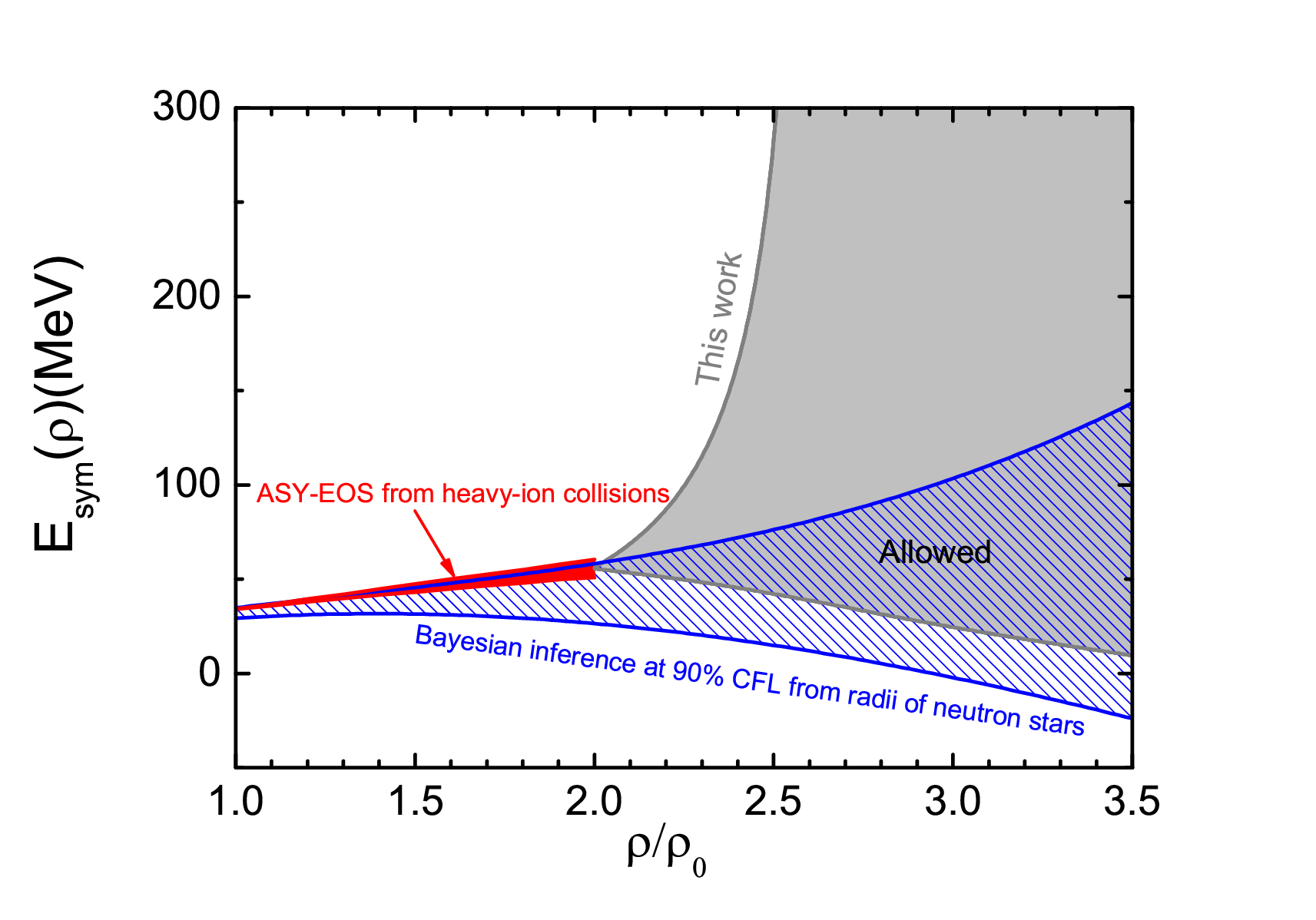}
  }
\caption{(Color online) The density dependence of nuclear symmetry energy at supra-saturation densities from the present work in comparison with that from the Bayesian inference of ref. \cite{Xie19} using the radius data of canonical neutron stars.}\label{figure4}
\end{center}
\end{figure}

Shown in Fig.4 is a comparison of the $E_{\rm{sym}}(\rho)$ from the present work with that from our recent Bayesian inference \cite{Xie19} from the radius data of canonical NSs.
The ASY-EOS band (red) is also shown. As discussed earlier, our integration of Eq. (\ref{esym}) started from $2\rho_0$ using the central value of the ASY-EOS band $E_{\rm{sym}}(\rho)(2\rho_0)=55.7$ MeV. It is interesting to see that while the two constraining bands from this work and our recent Bayesian analyses are both quite broad at densities above $2\rho_0$, they overlap. If both bands are equally trustable, their overlapping area provides a reasonably tight constraint on the symmetry energy of super-dense neutron-rich nuclear matter above $2\rho_0$. With respect to the current status of the field mentioned above, this constraining band for the $E_{\rm{sym}}(\rho)$ above $2\rho_0$ is a welcomed step forward. Testing and further reducing it using future observational/experimental data and/or other approaches will certainly be very useful. \\    

There are some caveats in our work reported here. Most importantly, as indicated in the first sentence of the abstract and discussed in detail in the text, our work is carried out within the minimum model of neutron stars consisting of only neutrons, protons and electrons. As such, our approach has some limitations and drawbacks. For example, various new particles and phases are predicted to occur above certain critical densities. In particular, muons are expected to first appear at suprasaturation densities. Within the $npe\mu$ model of neutron stars and under several approximations, it was estimated mostly analytically that the muon fraction reaches 1/16 which is half of the proton fraction when the baryon density goes to infinity or assuming muons are massless \cite{David}. In a more recent work involving one of us about Muonphilic Dark Matter in neutron stars within the $npe\mu$ model \cite{ZL-muon}, the maximum mass fraction of muons was found to be about 1\% on the causality surface when the symmetry energy is very stiff but quickly goes to zero when the symmetry energy is very soft favoring PNM in the core of neutron stars. Thus, a more reliable extraction from and studies of effects of high-density symmetry energy on neutron stars should consider muons. However, when other negatively charged particles are also considered simultaneously, such as the $\Delta^-$ that can appear above a critical density as low as $\rho_0$ depending on its completely undetermined coupling strength with the $\rho$-meson, see, e.g., refs. \cite{D1,Cai,AngLi,Armen1,Kim,Sahoo,Armen2}, the muon fraction decreases very quickly to zero at high densities. Thus, a more accurate study requires a much more comprehensive understanding about the composition and phases of dense matter beyond the abilities of both the $npe$ and $npe\mu$ models of neutron stars. Nevertheless, given all the uncertainties we discussed earlier and the fact that both of our input pressures were extracted from the terrestrial and astrophysical data under the assumption that there is no hadron-quark phase transition in dense matter, the minimum model for neutron stars is our best choice in this first application of the Eq. (\ref{esym}) without introducing more assumptions. As evidenced in many old and new studies in the literature, the minimum model provides invaluable guidances for our understanding of some properties of neutron stars and extracting some interesting underlying physics of dense matter. Moreover, as we emphasized earlier, our study is carried out in the context that the same data shown in Fig. 1 were used very recently in several publications by others in extracting the high-density symmetry energy within a sub-minimum model assuming neutron stars are made of purely neutrons. Keeping the simplicity and beauty of the minimum model of neutron stars, our study carried out here provides a useful guide for future more comprehensive studies of the high-density behavior of nuclear symmetry energy using more advanced models of neutron stars.\\

\noindent{\bf Summary:}
A new approach is proposed for inferring the proton fraction and symmetry energy of super-dense neutron-rich nucleonic matter directly from the barotropic
pressures $P_{\rm{NS}}(\rho)$ in NSs and $P_{\rm{SNM}}(\rho)$ for SNM from heavy-ion reactions. We have demonstrated that the approach is very efficient and viable. 
An application of this approach using the NS pressure from GW170817 and the SNM pressure from relativistic heavy-ion collisions provides a useful constraining band for the symmetry energy above twice the saturation density of nuclear matter. \\

\noindent{\bf Acknowledgement:} We thank Frank Hall for helpful discussions. Wen-Jie Xie is supported in part by the China Scholarship Council and appreciates the productive research conditions provided to him by Texas A\&M University-Commerce. BAL acknowledges the U.S. Department of Energy, Office of Science, under Award Number DE-SC0013702, the CUSTIPEN (China-U.S. Theory Institute for Physics with Exotic Nuclei) under the US Department of Energy Grant No. DE-SC0009971.
\\

\clearpage


\begin{thebibliography}{}

\bibitem{Tolman34} R. C. Tolman, Proc. Natl. Acad. Sci. U.S.A. 20 (1934) 3.
\bibitem{Oppenheimer39} J. Oppenheimer qnd G. Volkoff, Phys. Rev. 55 (1939) 374.

\bibitem{LIGO2018} B.P. Abbott \textit{et al.} (LIGO and Virgo Collaborations), Phys. Rev. Lett. \textbf{121} (2018) 161101.
\bibitem{Dan02} P. Danielewicz, R. Lacey, W.G. Lynch, Science \textbf{298} (2002) 1592.
\bibitem{Trau12} W. Trautmann, H.H. Wolter, Int. J. Mod. Phys. E \textbf{21} (2012) 1230003.
\bibitem{Garg18} U. Garg and G. Col\`{o}, Prog. Part. Nucl. Phys. \textbf{101} (2018) 55.
\bibitem{Wolfgang19} W. Trautmann, AIP Conference Proceedings 2127 (2019) 020003.

\bibitem{Bert}G. F. Bertsch and S. D. Gupta, Phys. Rep. {\bf 160} (1988) 189.

\bibitem{Sto86} H. St\"ocker, W. Greiner, Phys. Rep. {\bf 137} (1986) 277.

\bibitem{Cas90} W. Cassing, V. Metag, U. Mosel, K. Niita, Phys. Rep. {\bf 188} (1990) 363.

\bibitem{Bass} S.A. Bass et al., Prog. Part. Nucl. Phys. 41 (1998) 255.

\bibitem{Li2008} B.A. Li, L.W. Chen, C.M. Ko, Phys. Rep. \textbf{464} (2008) 113.

\bibitem{Xu19}J. Xu, Prog. Part. Nucl. Phys. 106 (2019) 312.

\bibitem{EPJA-review} Euro. Phys. J. A Topical Issue on ``First joint gravitational wave and electromagnetic observations: Implications for nuclear and particle physics", Eds. 
David Blaschke, Monica Colpi, Charles Horowitz and David Radice,
\url{https://link.springer.com/journal/10050/topicalCollection/AC_4be1cc6f5cdc3f8c4bf5cf69144c4b78}

\bibitem{JPG-review}Journal of Physics G: Nuclear and Particle Physics, Focused Issue on ``Hadrons \& Gravitational Waves After GW170817", Ed. Felipe J. Llanes-Estrada,
\url{https://iopscience.iop.org/journal/0954-3899/page/Hadrons_and_gravitational_waves_after_GW170817}

\bibitem{CUSTIPEN}
Proceedings of the Xiamen-CUSTIPEN Workshop on the Equation of State of Dense Neutron-Rich Matter in the Era of Gravitational Wave Astronomy, Eds. Ang Li, Bao-An Li and Furong Xu, AIP Conference Proceedings 2127 (2019) 010001. 
\url{https://aip.scitation.org/toc/apc/2127/1?expanded=2127}

\bibitem{ditoro} V. Baran, M. Colonna, V. Greco, M. Di Toro, Phys. Rep. \textbf{410} (2005) 335.
\bibitem{Steiner05} A.W. Steiner, M. Prakash, J.M. Lattimer, P.J. Ellis, Phys. Rep. \textbf{411} (2005) 325.
\bibitem{Lat12} J.M. Lattimer, Annu. Rev. Nucl. Part. Sci. \textbf{62} (2012) 485.
\bibitem{Tsang12} M.B. Tsang \textit{et al.}, Phys. Rev. C \textbf{86} (2012) 015803.

\bibitem{Chuck14} C.J. Horowitz et al., J. Phys. G \textbf{41} (2014) 093001.

\bibitem{Tesym} B.A. Li, \`{A}. Ramos, G. Verde, I. Vida\~{n}a (Eds.), \textit{Topical issue on nuclear symmetry energy}, Euro. Phys. J. A \textbf{50}, No.2 (2014).
\bibitem{Bal16} M. Baldo, G.F. Burgio, Prog. Part. Nucl. Phys. \textbf{91} (2016) 203.

\bibitem{Oertel17} M. Oertel, M. Hempel, T. Kl\"{a}hn, S. Typel, Rev. Mod. Phys. \textbf{89} (2017) 015007.

\bibitem{Chen2017} L.W. Chen, Nucl. Phys. Rev. \textbf{34} (2017) 20.

\bibitem{Li17} B.A. Li, Nuclear Physics News \textbf{27} (2017) 7.

\bibitem{Blas18} D. Blaschke, N. Chamel, {\it White Book of NewCompStar, European COST Action MP1304.}

\bibitem{PPNP-Li} B.A. Li, B.J. Cai, L.W. Chen, J. Xu, Prog. Part. Nucl. Phys. \textbf{99} (2018) 29.

\bibitem{MM1} J. Margueron, C.R. Hoffmann and F. Gulminelli, Phys. Rev. C97 (2018) 025805; {\it ibid} Phys. Rev. C97 (2018) 025806.

\bibitem{Baiotti}L. Baiotti, Prog. in Part. and Nucl. Phys. 109 (2019) 103714.

\bibitem{NBZ18} N.B. Zhang, B.A. Li and J. Xu, APJ, 859 (2018) 90.

\bibitem{NBZ19a} N.B. Zhang, B.A. Li, Euro. Phys. J. A \textbf{55} (2019) 39.

\bibitem{Zhang19apj} N.B. Zhang and B.A. Li,  APJ, 879 (2019) 99.

\bibitem{Tsang19}M. B. Tsang et al., Phys. Lett. {\bf B795} (2019) 533.

\bibitem{Bom91} I. Bombaci, U. Lombardo, Phys. Rev. C \textbf{44} (1991) 1892.

\bibitem{Lat01} J.M. Lattimer, M. Prakash, Astrophys. J. \textbf{550} (2001) 426.

\bibitem{Li2019review} B.A. Li, P.G. Krastev, D.H. Wen and N.B. Zhang, Euro. Phys. J. A 55 (2019) 117.

\bibitem{XiaoPRL} Z.G. Xiao, B.A. Li, L.W. Chen, G.C. Yong, M. Zhang, Phys. Rev. Lett. \textbf{102} (2009) 062502.

\bibitem{Xie19}W.J. Xie and B.A. Li,  APJ 883 (2019) 174.

\bibitem{Nakazato19} K. Nakazato and H. Suzuki, APJ 878 (2019) 25.

\bibitem{LWChen19} Y. Zhou, L.W. Chen and Z. Zhang, Phys. Rev. D 99 (2019) 121301.

\bibitem{PKU-Meng} H. Tong, P.W. Zhao and J. Meng, Phys. Rev. C 101 (2020) 035802.

\bibitem{Baillot19} N. Baillot d'Etivaux, S. Guillot, J. Margueron, N. A. Webb, M. Catelan and A. Reisenegger, APJ, 887 (2019) 48.

\bibitem{Zhou19} Y. Zhou and L.W. Chen, APJ, 886 (2019) 52.

\bibitem{Lynch09} W.G. Lynch et al., Prog. Nucl. Part. Phys. \textbf{62} (2009) 427.

\bibitem{russ11} P. Russotto \textit{et al.}, Phys. Lett. B \textbf{697} (2011) 471.

\bibitem{ASY-EOS} P. Russotto \textit{et al.}, Phys. Rev. C \textbf{94} (2016) 034608.

\bibitem{M217}H.T. Cromartie et~al., Nature Astronomy, 439 (2019).

\url{https://doi.org/10.1038/s41550-019-0880-2}

\bibitem{David} D. Blaschke, D.E. Alvarez-Castillo and T. Klahn, arxiv:1604.08575

\bibitem{ZL-muon}N.B. Zhang and B.A. Li, APJ 879 (2020) 99.

\bibitem{D1}A. Drago, A. Lavagno, G. Pagliara, and D. Pigato,  Phys. Rev. C {\bf 90} (2014) 065809.

\bibitem{Cai} B.J. Cai, F. J. Fattoyev, B.A.Li and W. G. Newton, 92 (2015) 015802.

\bibitem{AngLi} Z.Y. Zhu, A. Li, J.N. Hu and H. Sagawa, Phys. Rev. C {\bf 94}, 045803 (2016).

\bibitem{Armen1} J.J. Li and A. Sedrakian, Phys. Rev. C {\bf 100} (2019) 015809.

\bibitem{Kim} Y.Takeda, Y. Kim and M. Harada,  Phys. Rev. C {\bf 97} (2018) 065202.

\bibitem{Sahoo}H. S. Sahoo, G. Mitra, R. Mishra, P. K. Panda and B.A. Li, Phys. Rev. C {\bf 98} (2018) 045801.

\bibitem{Armen2}J.J. Li and A. Sedrakian, Astrophys. J. Lett, 874 (2019) L22


\end{thebibliography}
\end{document}